# Size-dependent concentration of $N^0$ paramagnetic centres in HPHT nanodiamonds

B.V. Yavkin *, G.V. Mamin, M.R. Gafurov, S.B. Orlinskii

Kazan Federal University, Kremlevskaya 18, 420008 Kazan, Russia

*E-mail: boris.yavkin@gmail.com

Size-calibrated commercial nanodiamonds synthesized by high-pressure high-temperature (HPHT) technique were studied by high-frequency W- and conventional X-band electron paramagnetic resonance (EPR) spectroscopy. The numbers of spins in the studied samples were estimated. The core-shell model of the HPHT nanodiamonds was proposed to explain the observed dependence of the concentration of the $N^0$ paramagnetic centers. Two other observed paramagnetic centers are attributed to the two types of structures in the nanodiamond shell.

## 1. Introduction

Nanodiamonds (NDs) have attracted increasing attention worldwide after a number of breakthroughs in synthesis, purification, isolation and their surface modification techniques achieved in the late 1990s [1]. At present, NDs can be synthesized at a relatively low cost by various synthesis techniques: detonation, laser ablation, high-energy ball milling of high-pressure high-temperature (HPHT) diamond microcrystals, plasma-assisted chemical vapour deposition (CVD), autoclave synthesis from supercritical fluids, chlorination of carbides, ion irradiation of graphite, electron irradiation of carbon onions [2]. Various applications of nanodiamonds have been suggested in recent years, particularly in the biological and medical fields, due to their excellent biocompatibility and superior optical properties [3-5]. The importance of nanodiamond surface study is also pumping by few vibrant research fields, like *in situ* magnetometry, thermometry and electric field measurements, quantum computing [6-8].

In as-produced detonation nanodiamonds the primary diamond particles are tightly bounded to each other forming large aggregates which are difficult to separate using conventional treatments [9]. New disaggregation approaches have been developed to obtain mono-dispersed nanodiamonds required by most biomedical applications. To date single-digital nanodiamonds with the size of about 4-5 nm in diameter can be achieved routinely and even smaller diamond particles are of interest [10]. The cause of strong aggregation of nanodiamond and the mechanism for the interaction between nanodiamonds and adsorbant are still challenges [11].

Samples produced using different synthesis techniques often have distinct sizes and surfaces which lead to significant differences in chemical reactivity and affinity to specific adsorbates. Purification treatments always introduce additional chemical terminations to the nanodiamond surface. Therefore, the characterisation of the ND nanoparticles and ND surfaces reported in the literature is often inconsistent [12].

Numerous and solid contributions to the study of nanodiamonds of various origin, size and surface termination were made by microscopic techniques such as transmission electron (TEM) [13], atomic force (AFM) [14] and confocal microscopy [15], NMR techniques [16], Raman spectroscopy [17], optically detected magnetic resonance (ODMR) [18, 19] technique. Unfortunately, the EPR experiments over the long period of time were on the backstage of nanodiamond research due to its limited sensitivity and lack of direct and unambiguous assignment of the observed signals [13]. In addition, despite large nitrogen content in nanodiamond, paramagnetic centers of nitrogen and nitrogen-vacancy centers that one could routinely observe in bulk diamond crystals [20-22] have not been observed in nanodiamond materials until recently [23-28].

In this work we investigate the evolutions of X-(9.6 GHz) and W-(94 GHz) bands EPR spectra in a series of the size-calibrated (10-1000 nm) commercial nanodiamonds prepared by the HPHT technique by exploiting conventional continuous wave (CW) methods of detection for qualitative and quantitative characterization of the core-shell structures of nanodiamonds.

## 2. Materials and Methods

A series of the size-calibrated nanodiamonds obtained from Tomei Diamond, Japan was studied: seven samples with the average size from 1 μm down to 10 nm, produced by HPHT synthesis from the diamond crystals of type Ib with the high nitrogen content in form of single substitutional nitrogen. Procedure of small nanodiamond particle production could be found elsewhere [9]. The sizes ($d$) of the investigated nanodiamonds and their notations throughout the article are presented in Table 1.

The EPR experimental data were acquired at room temperature by using the capabilities of Bruker ElexSys 580/680 spectrometer. CW EPR spectra in the W- band microwave frequency were used to identify the contributions from several paramagnetic centers to the overall EPR spectrum. To attain quantitative information about the paramagnetic centers of different nature, conventional X-band spectroscopy was employed and MgO:Mn crystal powder with the concentration of paramagnetic species of $5 \cdot 10^{14}$ spins per sample was used as a reference. To avoid saturation effects, the microwave power was chosen to be 0.2 μW for both X- and W- band measurements. The modeling of experimental EPR spectra was realized using EasySpin toolbox for Matlab [29].

**Table 1.** Size distribution of MD nanodiamond samples

| Sample | Median size $d$, nm |
|---|---|
| MD10 | 5-15 |
| MD20 | 15-25 |
| MD50 | 40-64 |
| MD100 | 90-129 |
| MD200 | 180-229 |
| MD500 | 480-529 |
| MD1000 | 980-1029 |

## 3. Results

The W-band EPR spectra and their simulations under the assumption of the presence of three paramagnetic centers with S = 1/2 (denoted as $N^0$, SC1 and SC2) are shown in Figure 1. The $g$-factors of two of them (SC1 and SC2) were calculated with respect to the reference of $g = 2.0024$ for $N^0$ center. The changes of $g$-factors with the MD size-were not monitored. The values of $g$-factors and linewidths for all of the samples and obtained paramagnetic centers are listed in Table 2.

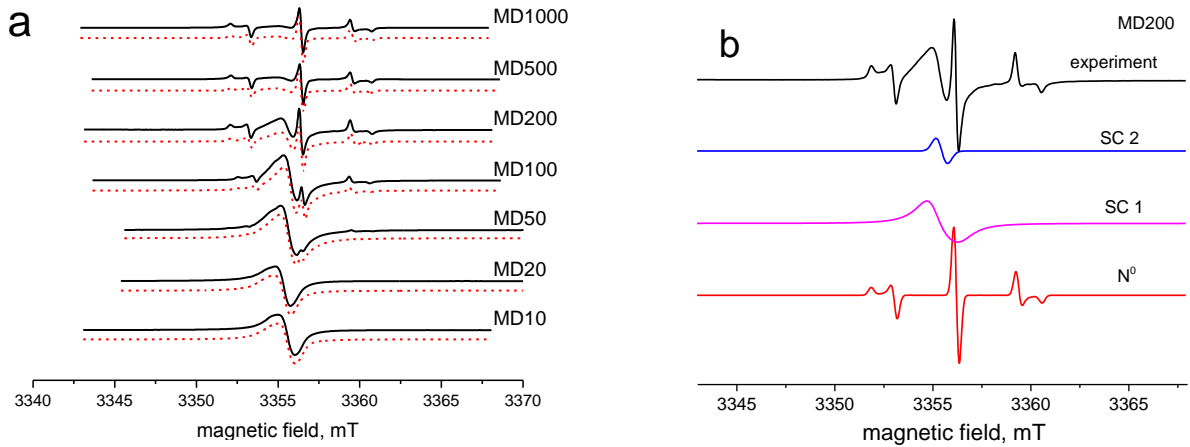

**Figure 1.** (Left panel) CW EPR spectra of nanodiamond samples (solid lines) and their corresponding simulations (red dashed lines). (Right panel) example of the spectral decomposition of the EPR spectrum for MD200 assuming the presence of three contributions. The spectra are detected at room temperature at 93.966 GHz.

The $N^0$ paramagnetic center shows up as a five lines powder EPR spectrum due to the anisotropic hyperfine interaction with $^{14}N$ of 99.6 % natural abundance and nuclear spin $I = 1$, described by the spin Hamiltonian $H = g\beta BS + A_\parallel S_Z I_Z + A_\perp (S_X I_X + S_Y I_Y)$ with $g$-factor of 2.0024 and hyperfine interaction constants $A_\perp = 2.92$ mT; $A_\parallel = 4.08$ mT. This description agrees both qualitatively and quantitatively with the previously published data on the nitrogen donors in the monocrystalline and powder diamond samples [20, 30]. Two other centers named SC1 and SC2 have $g$-factors of 2.0028(2) and 2.0029(2), respectively, and described by the lorentz and gauss lineshapes. The $g$-factors of SC1 and SC2 centers were determined relatively to the $g$-factor of $N^0$ center that was kept constant in our fittings for the whole size series; hyperfine interaction parameters and linewidth of nitrogen center were kept constant as well. It is interesting to note that the lineshape and width of SC1 component are practically coincide with those reported in one of the first study of deformation-induced paramagnetic centers in diamond [31, 32]. Separation of SC1 and SC2 centers became possible only by using high-frequency (high-field) EPR. One might try to describe both SC1 and SC2 as a single paramagnetic centre with the small anisotropy of $g$-factor, although this model would fail to explain the difference in the behaviour of signal intensities that correspond to the particular $g$-factor components with the nanoparticle size. Anisotropic g-factor model will lead to a constant intensities ratio between the signal components which is not the case for our experiment.

Quantitative study of number of paramagnetic centers was implemented by exploiting X- and W-band EPR measurements as follows. First, in X-band we have determined a total spin concentration in the sufficient amount of MD1000 (ca. 15 mg) and compared that with the spin concentration in the reference sample. Comparison was done by exploiting the double integration of the CW spectra. In order to keep the microwave tuning parameters unchanged we used only MD1000: the samples with the smaller sizes shifted the tuning conditions away from reference sample conditions. Second, we have re-calculated the extracted in this way the total spin concentration to the mass of the MD1000 used for the W-band experiments, which was determined to be about $3.2*10^{18}$ spin per gram. No changes in the tuning-matching conditions with the MD sample sizes were obtained in the W-band experiments that probably could be explained by the small amount of substances within the W-band tubes (less than 1 mg) and, therefore, the influence of the species on the microwave field distribution in the cavity is negligible. Third, we have extracted the relative concentrations of the observed paramagnetic centers in a series of size-calibrated samples using MD1000 as a reference for the W-band. The results of the numerical measurements are presented in Figures 2 and 3.

**Table 2.** Spectroscopic parameters of SC1, SC2 and $N^0$ centers obtained from the simulation of the W-band EPR spectra

| Sample | $g_{SC1}$ | lorentzian linewidth, mT | $g_{SC2}$ | gaussian linewidth, mT | $g_N^0$ | gaussian linewidth, mT | $A_\perp$, mT | $A_\parallel$, mT |
|---|---|---|---|---|---|---|---|---|
| MD10 | 2.0028(2) | 2.0(1) | 2.0029(2) | 0.8(1) | - | - | 2.92 | 4.08 |
| MD20 | 2.0028(2) | 1.9(1) | 2.0029(2) | 0.9(1) | - | - | 2.92 | 4.08 |
| MD50 | 2.0028(2) | 2.0(1) | 2.0029(2) | 0.7(1) | 2.0024 | 0.6 | 2.92 | 4.08 |
| MD100 | 2.0028(2) | 1.8(1) | 2.0029(2) | 0.7(1) | 2.0024 | 0.6 | 2.92 | 4.08 |
| MD200 | 2.0028(2) | 1.8(1) | 2.0029(2) | 0.7(1) | 2.0024 | 0.6 | 2.92 | 4.08 |
| MD500 | 2.0028(2) | 1.8(1) | 2.0029(2) | 0.7(1) | 2.0024 | 0.6 | 2.92 | 4.08 |
| MD1000 | 2.0028(2) | 1.8(1) | 2.0029(2) | 0.7(1) | 2.0024 | 0.6 | 2.92 | 4.08 |

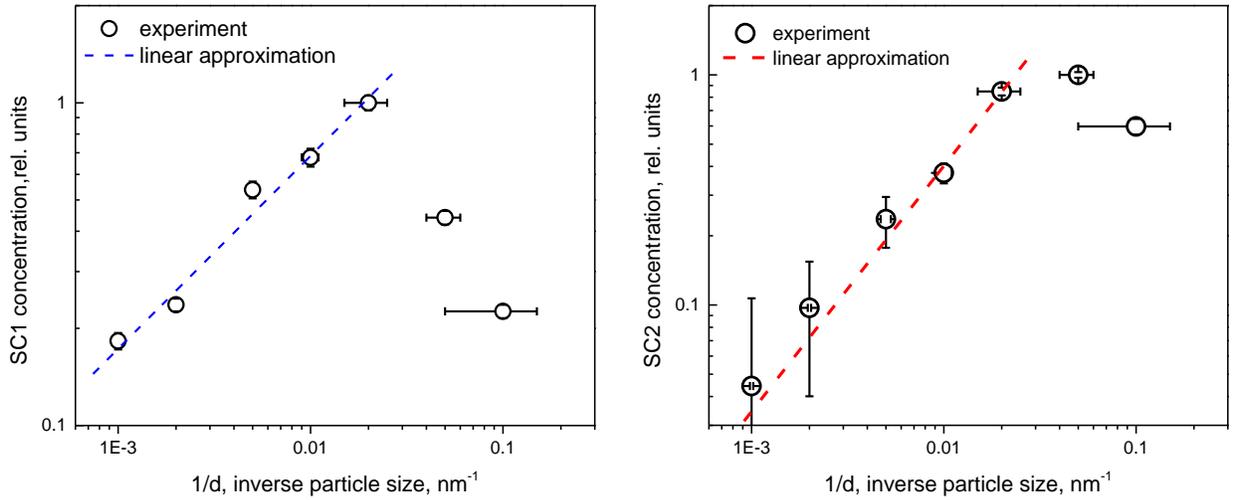

**Figure 2.** Relative concentrations of SC1 (left panel) and SC2 (right panel) paramagnetic centers as a function of inverse particle size along with their linear approximations for particle sizes larger than 20 nm.

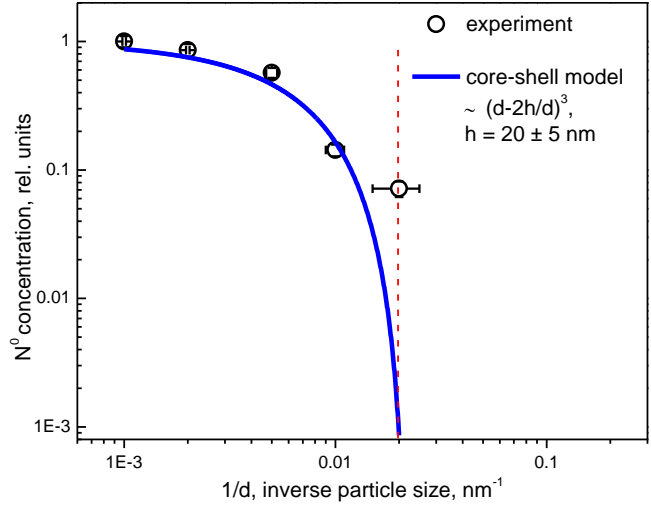

**Figure 3.** Relative concentrations of $N^0$ paramagnetic centers as a function of inverse particle size along with approximation for particles larger than 50 nm. Solid blue line is drawn according to the core-shell model approximation with $h = 20$ nm; red vertical dashed line corresponds to $d = 50$ nm.

## 4. Discussion

In order to explain the dependencies observed in Figure 2 and 3, one can propose a core-shell model of nanoparticle structure that is schematically shown in Figure 4. This is a relatively standard model for nanoparticle spectroscopy interpretation. For instance, it was used in joint XRD and EPR spectroscopy measurements of rare-earth doped fluorite nanoparticles studied in the work [33]. We suppose that $N^0$ centers are mainly concentrated in the core limited by the radius $R_C$ of the nanodiamond particle with the cross-section of diameter $d$ while in the surface layer of thickness $h$ other types of paramagnetic species prevail. The plots (figure 3, mainly) allow estimating the thickness as $h \approx 20(5)$ nm.

The lack of EPR signal from $N^0$ centers for MD20 and MD10 samples could be explained by either conversion of the neutrally charged nitrogen $N^0$ to paramagnetic-silent charge state or due to the nitrogen flushing outside of the diamond lattice region of the particle due to its energy-induced preferences [34]. Indeed, there are some reports for the nitrogen-vacancy (NV) charge conversion with the nanoparticle size [35, 36] or induced by surface chemical treatment [37] as well as by optical illumination [38]. Despite different electron spin density distribution for $N^0$ and NV, the certain similarity in their electronic structure and energy level positions within the band gap makes it possible to allow the analogous behavior with the nanoparticle size.

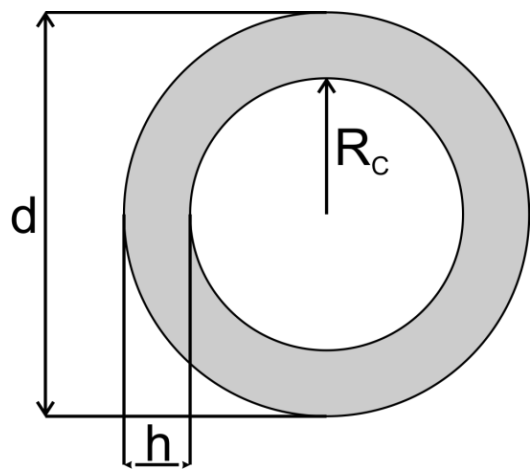

**Figure 4.** Schematic model of nanoparticle, where *d* is a cross section diameter (full size) of particle, $R_c$ is the diamond core radius, and *h* is the width of deformed diamond lattice region

The quantitative analysis of paramagnetic signals ascribed to surface layer(s) of particle is also of special interest. As stated above, the SC1 contribution was observed earlier in the study of deformed diamond crystals, but the SC2 component which is observed in this work was not detected directly in EPR spectra of nanodiamond before, except for instance the work of Nadolynny [24]. The EPR lineshape and width of SC1 center, along with almost perfect numerical agreement between total spin number of SC1 center and previously reported deformation-induced center in diamond [31] is probably a sufficient argument to relate this center to the mechanically-deformed paramagnetic species that were observed in diamonds previously. Linear dependence of the SC1 concentration with the inverse particle size at d ≥ 50 nm (Figure 2) confirms homogeneous distribution of damage-induced paramagnetic centers over the geometrically distinct shell with the width *h*. The same behavior is to observe for SC2 centre (Figure 2). However, the MD10 and MD20 samples exhibit deviation of the SC1 and SC2 size-dependent concentration from the linear law, namely the concentration of SC2 is independent on the particle size while the EPR intensity for SC1 centre shows a sharp decrease. The origin of this decrement is not discussed in this work, but it is clear that the presented core-shell model is not an adequate one for the particles below 50 nm.

## 5. Conclusion
For the first time, based on the combination of conventional X-band and high-frequency (W-band) EPR measurements, the core-shell model of the HPHT-synthesized nanodiamond particle is proposed and shown to be in consistency with the numerous Raman, optical and NMR studies. Apparently, the nanoparticle consists of three distinct structural layers characterized by the specific paramagnetic centers. Nitrogen $N^0$ centers are located within the diamond core of nanoparticle, SC1 paramagnetic center are filled the inner surface region with the thickness of about 20 nm of the deformed diamond lattice while SC2 centers are localized in the outer shell of mainly non-diamond carbon atoms.

The spin-Hamiltonian parameters for the SC1 center are in qualitative agreement with previously published papers devoted to the studies of the deformed synthetic diamond with the sizes from hundreds of micrometers down to 1 μm. SC2 center exhibits purely surface behavior and its concentration is inversely proportional, in large extent, to nanoparticle size. It is shown that the amount of the paramagnetic nitrogen centers becomes negligible at HPHT diamonds dimensions of less than approx. 40-50 nm probably due to their conversion to the nonparamagnetic state.

**Acknowledgements**

Authors are grateful to Igor Vlasov from GPI RAS for providing MD samples for research. This work was partially funded by the subsidy allocated to Kazan Federal University for the project part of the state assignment in the sphere of scientific activities.